\documentstyle[aps]{revtex}
\def \be   {\begin{equation}}
\def \ee   {\end{equation}}
\def \l {\label}
\begin{document}
\input epsf
\baselineskip=25pt
\title{Discrete gauge fields}
\author{Manoelito M de Souza}
\address{Universidade Federal do Esp\'{\i}rito Santo - Departamento de
F\'{\i}sica\\29065.900 -Vit\'oria-ES-Brasil}
\date{\today}
\maketitle
\begin{abstract}
\noindent The use of proper time as a tool for causality implementation in field theory is clarified and extended to allow a manifestly covariant definition of discrete fields proper to be applied in field theory and quantum mechanics.  It implies on a constraint between a radiation field and its sources, valid in principle for all fundamental interactions and with a solid experimental confirmation for the electromagnetic one.  Some results of its applications to an abelian classical theory (electrodynamics taken as a first example), and with the discrete field being regarded as a classical representation of the field quantum (photon) are anticipated in order to illuminate the physical meaning and the origins of gauge fields and of their symmetries and singularities. They are associated to a loss of field-source coherence.
\end{abstract}
\begin{center}
\end{center}
\section{Introduction}
In this paper we introduce just one basic thing with respect to the usual formulation of field theory: we replace the lightcone by the lightcone generator, a straight line embedded in a $(3+1)$ Minkowski spacetime, as the support (or domain) for the definition of radiation fields. Everything else here is a consequence of this single modification in the standard field theory. 
Just as a heuristic motivation, let us consider an spherical electromagnetic signal emitted from a point P, for fixing the idea.  It propagates on the P-light-cone as a wave, distributed over an ever expanding sphere centred at P. It is a travelling spherical wave, an example of a field $A(x)$ defined with  support on the lightcone. The field intensity decreases with the square of the distance from the origin and, consequently, we have problems with an infinite energy-momentum density if we trace back this wave to P, its starting point where, let's say, we find a point electron. According to quantum physics this continuity in the classical field is just apparent as this wave front is a collection of a, huge but finite, number of  field quanta. Let us imagine the wave front as a set of moving points, each one on a lightcone generator. In this new discrete approach to field theory the classical spherical wave front will be seen as just an effective representation of a collection of point fields, to be properly defined below in a covariant way. 

We call discrete field a field defined on such a restricted support for a distinction from the standard extended one. We do it in a more generic way valid for massive and massless fields, and although we actually apply it here just to the Maxwell's theory it can readily be extended to non-linear and to non-abelian field theories. This explains the generic reference to gauge fields in the title. 

This approach came from a study of classical electromagnetic radiation on its limiting of zero distance from its sources \cite{hep-th/9610028} where it was shown that although the potential field has support on the lightcone (which is the origin of most of its problems), its force field (the Maxwell's stress tensor) has, effectively, support on the lightcone generator and that this knowledge is sufficient for solving old problems of infinite energy and causality violation \cite{Rorhlich,Jackson}. The enormous simplification power of this discrete-field concept is exhibited in an application to the general theory of relativity \cite{gr-qc/9801040}: the highly non-linear field equations are reduced, without any approximation, to the wave equation in a (3+1) Minkowski spacetime, and yet from its discrete solutions one can, in principle, retrieve any extended solution, according to the property 12 in the list below.

This simple change of support has two immediate dynamical consequences. The first one is expressed as a constraint between the source's acceleration and the direction of propagation of the field: in the instantaneous rest frame of the accelerated source, the field is always emitted along a direction orthogonal to its source acceleration. For the electromagnetic interaction, in particular, this constraint has a solid experimental confirmation that validates its implementation as a basic physical input. But since it was derived on very generic grounds without reference to any specific interaction it is expected to be valid for all fundamental interactions. This relation, eq. (\ref{dA0}) of Section III, is the paper most important single result and the basic cause behind all the others. 

The second consequence is that any discrete field is associated to a continuity equation (as the field is restricted to be a flux on a line) which, with the above causal constraint, leads to an anticipated satisfaction of certain restriction that  are expressed either as conservation laws (of charge, energy, momentum, etc) or as a covariant gauge condition, for example, that for a discrete field is reduced to an identity.
 
We find the discrete solutions to the standard wave equation in $(3+1)$ dimensions to show that, besides satisfying the above mentioned causal constraint, they have the following rather extensive set of remarkable properties:
\begin{enumerate}
\item The definition of a discrete field is consistent with the wave equation in the sense that it is sustained during the field propagation;
\item They are finite, propagating and point-like fields as they are different of zero only at a single point at a time. Each discrete field represents just a single point in phase space;
\item They are fields without singularity, in the sense of infinity. They propagate without changing their amplitudes;
\item There are no advanced discrete field; they are all retarded fields and their interactions with their sources are similar to processes of creation and annihilation of particles;
\item The homogeneous wave equation has only trivially constant solutions. The discrete field is univocally determined by the state of motion of its source, so it has no gauge freedom. It is not a gauge field;
\item There is a one-to-one map, a correlation, between each discrete field and its source at its retarded time (its creation event);
\item An anti-symmetric force field is a necessary consequence of causality and Lorentz covariance;
\item They are transversal fields, they have no spurious (non physical) degrees of freedom. The Lorentz gauge condition for them is just an identity.
\item Each discrete field has well defined and everywhere conserved energy and momentum;
\item The wave equation has just one kind of discrete solution although they may have different contents of energy and momentum; 
\item Although discrete in space they produce interference effects just like the extended field; 
\item The smearing of a discrete field on the lightcone, with the neglect of the causal constraint, reproduces the standard extended field. Any extended field can be seen as an average effective field from a linear combination of discrete fields;
\item The obtention of extended fields from the discrete ones, although being a mathematically well defined operation, implies on a loss of information (the causal constraint) so that its inverse  is, in general, not possible unless the lost information be included back.
\end{enumerate}
With respect to this list of properties there are two considerations that must be made. First, we can conclude that the standard and the discrete fields do not share the same properties. Studying the passage from the discrete to the extended field one can have a deeper understanding of the physical meaning and origin of many of the properties and problems of the extended fields. Most of them are consequences of their lightcone support. The singularity, for example, is just a reflex of the lightcone vertex. The singularity and the divergent self-field energy of the Li\`enard-Wiechert solution of classical electrodynamics are not consequences of a point electron but of their lightcone support\cite{hep-th/9610028}. The same goes for the singularities in general relativity \cite{gr-qc/9801040} and for the problems with the Lorentz-Dirac equation \cite{hep-th/9610028}. The gauge symmetry of the standard field is a consequence of the smearing process that erases property 9, the correlation between a discrete field and its source at its retarded time.

The second consideration is about the meaning of the very discrete field. The properties listed above show that it must be regarded as a physical entity by itself and not just a mathematical artifact. After all, it carries conserved and well defined energy and momentum. It is an elementary unit field that enters in the constitution of any extended field. For the electromagnetic field, which is the case considered in this paper, there is no alternative but to associate it, in some way to be made more precise in the future, to the photon since this is the only observed electromagnetic elementary manifestation.  We know of course that this is not so simple, considering that a discrete field does not completely corresponds to our present image of the photon as a quantum or dual object. But the real point here, we believe, is that although we have learned how to work with quantum objects, how to extract meaningful results from their quantum theories we still do not know what they really are. On dealing with a discrete electromagnetic field we will pragmatically refer to it as the ``classical photon" and leave any further discussion about it for a future work.

The following analogy may be very elucidative of our viewpoint. The relationship between a discrete field and the standard extended field is similar to the one between thermodynamics and statistical mechanics. The meaning and the origins of thermodynamic variables (P, T, S, etc) are illuminated by statistical mechanics from the knowledge of basic structural elements unknown to thermodynamics. Similarly, properties of an extended field (its problems, symmetries and singularities) are better understood from its relation to its discrete field. Thermodynamics is an effective description in terms of average valued properties of the more basic structures, the molecules,  considered in statistical mechanics; likewise, a field theory (general relativity and quantum mechanics included) are retrieved from their respective formulation in terms of discrete fields as effective average-valued descriptions. Statistical mechanics does not change the status of any of the four laws of thermodynamics, it just put them in a deeper perspective. Analogously, nothing changes in the validity and in the fundamental character of the Maxwell's equations with the discrete fields and they are equally seen from a deeper perspective. Although not being a fundamental theory thermodynamics is successfully used when this is more convenient than using the known basic molecular structures. The same goes for the extended and the discrete fields.

This paper is organized in the following way. In Section II we discuss some new and old problems of field theory with the objective of highlighting some questions to be fully appreciated after the introduction of discrete fields. In Section III a new geometrical way, based on proper time, of implementing relativistic causality is introduced. It extends the concept of local causality to make possible the definition of discrete fields. Its implications to the field dynamics is exhibited in the eq. (\ref{dA0}). Discrete field are defined in Section IV. In Section V we find and discuss the discrete Green's functions, and in Section VI, how to retrieve from them the extended (standard) ones.  The relationship between the discrete and the extended field formalisms is discussed in Section VII. The photon as a discrete field is seen in Section VIII. In Section IX  we discuss the surging of singularity, gauge freedom and duality. The Section X deals with the angular dependence in the emission of discrete fields. The paper ends with some final comments and the conclusions, Section XI. 

\section{Some old and new problems with fields}
This paper is about radiation fields, their meaning, origins and properties. It is concerned with some fitting problems in their mathematical descriptions. Some are old ones and have been discussed since the early conceptions of field theory but others seemingly have not yet been completely perceived as such. They are closely related to the new discrete field approach that is being presented here. 

To a solution from an inhomogeneous equation one can always add any solution from the homogeneous. This leaves lose the connection between a field and its source which requires the imposition of boundary conditions to be fixed. There is no problem with this as far as just mathematics is concerned, but it is not so if it is applied to a radiation field , for example. Physically this cannot be satisfactory because what is locally produced should not be affected by what will become of it at large. A source should always react in the same way under a same stimulus regardless any boundary conditions; it shouldn't know anything about them.  So, the source's state of movement should be enough for completely determining its field. The well known fact that boundary conditions are required for that constitutes a logical conflict from the physical viewpoint, a paradox. This is a fitting problem between radiation and the mathematics used in its description. 

While there is a clear mathematical distinction between solutions from an inhomogeneous equation and from  its homogeneous one, from a physicist's viewpoint this is very annoying when these solutions describe radiation fields because then there should not exist any non trivial difference between them.
There is no radiation field without a source\footnote{This could be seen as just an assertion but it finds strong support with the discrete field concept in Section V}; it does not spontaneously outgrow from the empty. A radiation field is  a solution of an inhomogeneous equation at the moment of its generation only, afterwards as it propagates away from its sources it becomes a solution of its homogeneous equation. They are and must be, of course, the same solution. 
To have only a homogeneous solution, that is a solution that never satisfies the inhomogeneous equation, means not having the complete information. Something is missing, for example, its birth certificate, i.e. the information of when and where it was generated. A more appropriate mathematical description of radiation fields should not allow so much distinction between solutions from the homogeneous and from the inhomogeneous equations, and such a lose connection between a field and its sources. After all, these mathematical distinctions do not correspond to what is observed: there are not two classes of, let's say, photons associated respectively to the homogeneous and to the inhomogeneous equations. It would be contradictory if a photon that is not generated from a source interaction could, by pair creation, generate new sources.

These could be taken just as subtle details if there were not many others important, but apparently unrelated, examples of conflict between a physical field and its adopted mathematical  representation. They are all consequences of the concept of field as ``something" non-localized, continuously extended over space. Presently, perhaps, the most important of such questions for the physicists be their difficulty on understanding the wave-particle duality of quantum mechanics. It is particularly interesting here for exposing completely this kind of conflict: the field is welcomed for explaining the wave-like properties of radiation and for allowing their extension to matter but it becomes an inconvenience when one stumbles with unsurmountable problems for explaining the particle-like properties. The well known problem here is of collapsing an extended wave function. The discrete field carries both properties: the wave-like property one, for being a field, and the particle-like for being discrete. It is an embodiment of the wave-particle duality and it allows a better understanding of the meaning of quantum mechanics although we will have just a glance over these questions here. 

No mention to old and well known problems with infinities that plague both classical \cite{Rorhlich} and quantum \cite{Muta} field theories; all consequences of extended fields. That a classical point electron\footnote{A relativistic extended one would just bring in more problems \cite{Rorhlich,Dixon}} has an infinite energy stored on its self-field, that it generates, when accelerated, radiation with an initially infinite density of energy, and that its equation of motion violates causality are all consequences \cite{hep-th/9610028} of the concept of field as an extended object. That a mensurable physical property at a certain moment of time assumes an infinite value points to an inadequacy of the mathematical formalism.  It is true that nowadays one is so used to the pragmatic efficiency and success of the renormalization program in quantum field theory that many are not afraid anymore of infinities. But infinities are inherent to a formalism based on fields with an infinite number of degrees of freedom and they are always an indication of an improper formalism.

\begin{center} Gauge fields
\end{center}

It is well known that a constraint imposed on a field may reduce the number of its independent components but  the origin and meaning of gauge freedom was not completely clear. This paper hopefully will help on making it clearer. 
The questions we want to discuss here are pertinent to all gauge fields, quantum or classical, abelian or non-abelian, but for obvious reasons of simplicity we will take the Maxwell theory as a  case on studying.

The Maxwell's equations 
\be
\l{hM}
\varepsilon^{\mu\nu\rho\sigma}\partial_{\nu}F_{\rho\sigma}=0,
\ee
\be
\l{iM}
\partial_{\nu}F_{\mu\nu}= J^{\mu},
\ee
contain all the accumulated knowledge about the electromagnetic field $F$ and its interactions  with the sources $J$. The homogeneous set of equations has 
\be
\l{hs}
F_{\mu\nu}=\partial_{\nu}A_{\mu}-\partial_{\mu}A_{\nu},
\ee
as a generic solution that turns the second set of equations into
\be
\l{iMa}
\Box A^{\mu}-\partial^{\mu}\partial.A= J^{\mu},
\ee
which is not solvable because the four equations cannot be disentangled. But only $F$, and not $A$, is directly  accessible to experimental detection, and $F$ is an antisymmetric tensor. All this leads to the notions of gauge fields and of gauge freedom in classical electrodynamics
\be
\l{gf}
\cases{A\Rightarrow A+\partial\Lambda,&\cr
	F\Rightarrow F.&\cr}
\ee
So we can impose the gauge condition
\be
\l{L}
\partial.A=0,
\ee  
which is, actually, an integrability condition for (\ref{iMa}) as it is then reduced to four uncoupled equations
\be
\l{we}
\Box A=J.
\ee
The wave equation is indeed a basic and universal field equation in physics. A well known generalization of this simple standard approach is common to all gauge fields but there are some lingering questions. It is apparent, for example, from (\ref{hM}) and (\ref{iM}) and it is also valid for their non abelian generalizations, that gauge freedom and charge conservation would be consequences of the anti-symmetry of the interaction field; what is the reason that makes anti-symmetric all fundamental interaction fields? Or, more than that, what makes all of them be gauge fields? The non-abelian analogues of $F$, although not gauge invariant and with a larger internal symmetry, are anti-symmetric in their lorentzian indices. In general relativity the potential, but not the field is described by a symmetric tensor. It will be shown in Section VII that charge conservation is not a consequence of gauge freedom and that both, charge conservation and gauge freedom, are not consequences of the field anti-symmetry  (although it certainly would not be possible to have gauge freedom without introducing an anti-symmetry somewhere);  this anti-symmetry and charge conservation are necessary consequences of causality and Lorentz covariance only; gauge freedom, like the problem of lose connection between a field and its  source, comes from the field extended representation. These results come from a discrete-field approach where the discussion above about the integrability of Maxwell's equation could be presented in a reversed order:  starting from the wave equation (\ref{wef}) below, to obtain all the above equations (\ref{hM},\ref{iM},\ref{hs},\ref{iMa},\ref{gf},\ref{L}, and \ref{we}) without any further assumption. The field described by equation (\ref{wef}) must be then more fundamental and defined in a background  endowed with more physical information than their respective ones in the standard formalism as they dispense extra assumptions (eqs. \ref{hs} and \ref{L}) in order to rescue the original equations (\ref{hM}) and (\ref{iM}). It must, however, be left clear that this is not in anyway a derivation of Maxwell's equations from more fundamental principles as they, like the Newton's equations, the Schrodinger's equation and many others, represent new axioms. The physical content of eqs. (3) and (6) is already implicit in the background necessary for defining the new discrete field.

\section{ Local and extended causality}

\noindent We recur to causality for defining the discrete field in a consistent and manifestly covariant way. Any given pair of events on Minkowski spacetime defines a four-vector $\Delta x.$ If a  $\Delta x$ is connected to the propagation of a free physical object (a signal, a particle, a field, etc) it is constrained to satisfy
\be
\label{1}
\Delta\tau^2=-\Delta x^{2}, 
\ee 
where $\tau$ is a real-valued parameter. We use a metric $\eta=diag(1,1,1,-1)$. $\Delta\tau$ is the invariant length or norm of $\Delta x$. So, the constraint (\ref{1}) just expresses that $\Delta x$ cannot be spacelike. A physical object does not propagate over a spacelike $\Delta x.$ This is {\it local causality}.  Geometrically it is the definition of a three-dimensional double hypercone; $\Delta x$ is the four-vector separation between a generic event $x^{\mu}\equiv({\vec x},t)$ and the hypercone vertex. See the Figure 1.

\vglue-2cm 
\begin{minipage}[]{5.0cm}\hglue-3.50cm
\parbox[]{5.0cm}{
\begin{figure}
\epsfxsize=400pt
\epsfbox{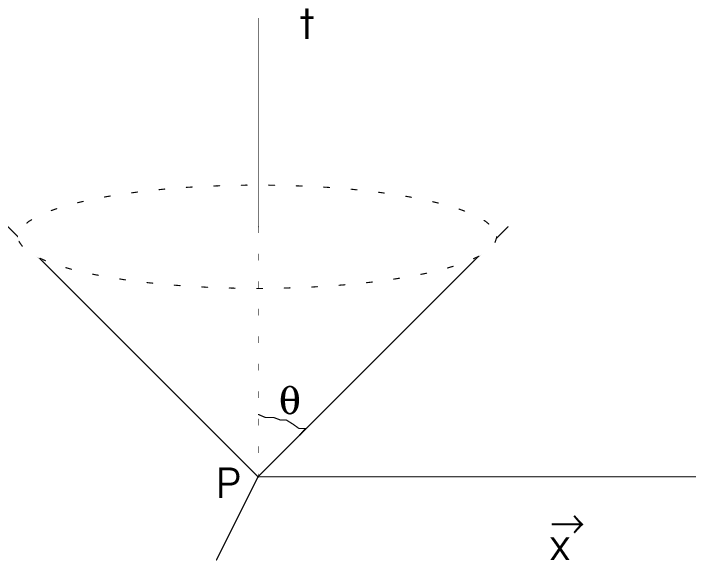}
\end{figure}}
\end{minipage}\hfill
\vglue-4cm
\vglue-12cm
\hglue7.0cm
\begin{minipage}[]{7.0cm}\hglue7.0cm
\begin{figure}
\l{f1}
\parbox[b]{7.0cm}
{\vglue-7cm \parbox[t]{7.0cm}{\vglue-1cm
\caption[Local causality]{The relation $\Delta\tau^2=-\Delta x^{2},$ a causality constraint, is seen as a restriction of access to regions of spacetime. It defines a three-dimension hypercone which is the spacetime available to a free physical object at the hypercone vertex. The object is constrained to be on the hypercone.}}}
\end{figure}
\end{minipage}\\ \mbox{}

The variation of proper time is defined through intervals associated to the propagation of a free field in the Minkowski spacetime and not through its association to trajectories as it is  usual. This subtlety avoids a restriction to classical contexts and allows its application to quantum physics too.

The validity of eq. (\ref{1}) is conditioned to its application to a free field but this is not such a great limitation as it seems to be at a first look if we assume that at a fundamental level all interactions are discrete. It describes the free evolution of an interacting field between any two consecutive interaction events. 

Local causality is usually implemented in special relativity through the use of lightcones by requiring that massive and massless objects remain, respectively inside and outside, a lightcone.  Our way of implementing the same relativistic causality is of using hypercones (not necessarily lightcones) even for massive physical objects (the expression physical object is used here for not distinguishing between particles and fields) as a constraint on their propagation. 
In spacetime a field is defined on hypersurfaces: hyperplanes for newtonian fields, for example, and hypercones for relativistic free fields. Think of an wave front, for example, and think of it as a continuous set of moving points, then each point of it is on a worldline tangent to a generator of its instantaneous hypercone. 

 This conic hypersurface, in field theory, is the support for the propagation of a free field: it cannot be inside nor outside but only on the hypercone. The hypercone-aperture angle $\theta$ is given by $\tan\theta=\frac{|\Delta {\vec x}|}{|\Delta t|},\; c=1,$ or equivalently, $\Delta\tau^{2}=(\Delta t)^{2}(1-\tan^{2}\theta).$  The speed of propagation determines the hypercone aperture (rapidity). A change of the supporting hypercone corresponds to a change of speed of propagation and is an indication of interaction.
Special Relativity restricts $\theta$ to the range $0\le\theta\le\frac{\pi}{4},$ which corresponds to a restriction on $\Delta\tau:$ $0\le|\Delta\tau|\le|\Delta t|.$ The lightcone ($\theta=\frac{\pi}{4},$ or $|\Delta\tau|=0$) and the t-axis in the object rest-frame ($\theta=0,$ or $|\Delta\tau|=|\Delta t|$) are the extremal cases. The lightcone for objects with the speed of light and a parallel line to the time-axis for each point of an static field or of a massive object on its rest frame. 

For defining a discrete field we will need a more restrictive constraint:
\be
\l{f}
\Delta\tau+  f.\Delta x=0,
\ee
where $f$ is defined by $f^{\mu}=\frac{dx^{\mu}}{d\tau},$ a constant  four-vector tangent to the hypercone; it is  timelike $(f^{2}=-1$) if $\Delta\tau\ne0,$  or lightlike $(f^{2}=0$) if $\Delta\tau=0$. \\ It implies that
\be
\l{fmu}
f_{\mu}=-\frac{\partial\tau}{\partial x^{\mu}}.
\ee 
The equation (\ref{f}) defines a hyperplane tangent to the hypercone (\ref{1}). Together, eqs. (\ref{1}) and (\ref{f}) define a hypercone generator $f$, tangent to $f^{\mu}$. A fixed four-vector $f^{\mu}$ at a point labels a fibre in the spacetime, a straight line tangent to $f^{\mu}$, the $f$-generator of the local hypercone (\ref{1}).\\
Extended causality is the imposition of both eqs. (\ref{1}) and (\ref{f}) to the propagation of a  physical object. Geometrically, it is a requirement that the point object remains on the hypercone generator $f$. This corresponds to a change in our perception of the spacetime causal structure; instead of seeing it as a local foliation of hypercones (\ref{1}) we see it as  congruences of lines (eqs. \ref{1} and \ref{f} together); instead of dealing with continuous and extended objects, like a field for example, we treat them  as sets of points. 

\begin{center}
Causality and dynamics
\end{center}
In general eqs. (\ref{1}) and (\ref{f}) are just two kinematical constraints on the field propagation but the constraint (\ref{f}) has a very important dynamical content when $\Delta x$ describes also the  spacetime separation between two physical objects like a source and its field as we discuss now.  
$\Delta x=x-z(\tau)$ for a field emitted by a point charge at $z(\tau)$, which is taken, using (\ref{1}), as parameterized by  its $\tau$.  There is a crux difference in (\ref{f}) in the limiting of $x$ tending to $z(\tau)$ as there is then a singular change in the character of $f$ that is essential for understanding classical electrodynamics in this limit \cite{hep-th/9610028}.

For a massless field the restriction (\ref{f}) is reduced to $f.(x-z(\tau))=0$ and this implies that the event $x$, where the field is being observed, and the charge retarded position  $z(\tau)$ must belong to a same straight line $f$.  It is not necessary to explicitly distinguish a generic $\tau$ from a $\tau$ at a retarded position, as the situations considered in this note, from now on, will always refer to the last one. 

More information can be extracted from this constraint as $\partial_{\mu}f.(x-z){\Big |}_{f}=0$ implies on 
\be
\l{fv} 
f.V{\Big |}_{f}=-1,
\ee
where $V=({\vec V},V_{4})=\frac{dz}{d\tau}$. This relation  may be seen as a covariant normalization of the time component of $f$ to 1 in the charge rest-frame at its retarded time, 
\be
\l{f4}
f^{4}{\Big |}_{{f\;\;}\atop{{{\vec V}=0}}}=|{\vec f}|{\Big |}_{{f\;\;}\atop{{\vec V}=0}}=1.
\ee
With another derivation and with $a^{\mu}=\frac{dV^{\mu}}{d\tau}$ we get from (\ref{fv}) the single most important equation of this paper:
\be
\l{dA0}
a.f{\Big |}_{f}=0,
\ee
a constraint 
between the direction $f$ along which the signal is emitted (absorbed) and the instantaneous change in the charge state of motion at the retarded (advanced) time. It implies that
\be
\l{a4}
a_{4}=\frac{{\vec a}.{\vec f}}{f_{4}}{\Big|}_{f},
\ee
whereas $a.V\equiv0$ leads to
$a_{4}=\frac{{\vec a}.{\vec V}}{V_{4}}{\Big|}_{f},$
and so we have that in the charge instantaneous rest frame at the emission (absorption) time ${\vec a}$ and ${\vec f}$ are orthogonal vectors,
\be
\l{af}
{\vec a}.{\vec f}{\Big |}_{{f\;\;}\atop{{\vec V}=0}}=0.
\ee
For the electromagnetic field this is an old well known and experimentally confirmed fact but it takes  the whole apparatus of Maxwell's theory in order to be demonstrated \cite{Rorhlichp112}. The well-known \cite{Jackson} distribution of emitted electromagnetic radiation from accelerated charges can readily be derived (Section VIII). Its experimental confirmation is a direct validation of extended causality. The constraint (\ref{dA0}) has been obtained here on very generic grounds of causality, without reference to any specific interaction, which makes of it a universal relation, valid for all kinds of fields and sources. This is a most remarkable point, that this same behaviour is predicted to hold for all fundamental (strong, weak, electromagnetic and gravitational) interactions. It will be shown in the following that eq. (\ref{dA0}) is the basic cause behind other important results like conservation laws (of charge, momentum, energy, etc), for example. 
\section{Discrete fields}
\noindent We turn now to the question of how to define, in a consistent and manifestly covariant way, a field with support on a generic fibre $f$, a $(1+1)$-manifold embedded on a $(3+1)$-Minkowski spacetime. According to Einstein's special relativity to any physical object is associated a particular flow of time given by  its  proper time.  
All  fields are, in this way, tied to the proper time of their (massive or not) point source, or better, to their birth event. There is nothing new or special on this: a massless field, for example, propagates on the lightcone and so its proper time do not change; its clock keeps marking the time of its creation, the instantaneous proper-time of its source at the event of its emission. The other extreme situation is of an static field for which  $\Delta\tau=\Delta t$ at each point. 
We use this here as a way of implementing causality \cite{hep-th/9610028,gr-qc/9801040,BJP}. With $\tau$  being a known function of $x$, a solution of eq. (\ref{1}),  
\be
\tau=\tau_{0}\pm\sqrt{-(\Delta x)^{2}},
\ee
to write $A(x)$ for a field is the same (or almost) as writing $A(x,\tau(x)).$ The subtlety of including $\tau(x)$ is of encoding in the very field the causal constraint (\ref{1}) on its propagation. 
Then to say, for example, that $\Delta\tau=0$ for a field, is just an implicit way of requiring that it propagates on a lightcone.  In order to implement local causality for a given field $A(x)$, therefore, we just have to insert in it an explicit dependence on $\tau(x)$. For example, the replacement
$$A(x)\Longrightarrow A(x-z,\tau(x)-\tau(z)){\Big |}_{{\tau(x)=\tau(z)}}$$ describes a radiation field on the lightcone propagating from an event $z$ to an event $x$. We will write just  $A(x)\Longrightarrow A(x,\tau){\Big |}_{\Delta\tau=0}$, for short. An static field, following this line of thought, is just a particular case where  $$A(x)\Longrightarrow A(x,\tau){\Big |}_{{\Delta\tau=\Delta t}}$$ which is, naturally, a frame dependent expression, in contradistinction to the previous one.  
 
The Lienard-Wiechert solution \cite{Rorhlich,Teitelboim} is a well known example of how a propagating field depends on the proper time of its source. This dependence, let us repeat, is just a form of causality implementation. The distinctive subtlety introduced here is of considering the $\Delta\tau$ of each $\Delta x$ associated to the propagation of any physical object; one for each point source and one for each field. The use of $\tau$ is very frequent in the literature (see \cite{tau} and the references therein), always followed by some qualification like the invariant, the universal,  the historic and so many others for time, and with various interpretations and goals.  Here $\Delta\tau$ is always the length of a $\Delta x$ and is always introduced in association to this vision of causality.

Whereas the implementation of local causality requires a field defined with support on hypercones,  for extended causality it is required a field with support on a line $f$:
$$A(x)\Longrightarrow A(x-z,\tau(x)-\tau(z)){\Big |}_{{\Delta\tau+f.\Delta x=0}\atop{\Delta\tau^2+\Delta x^2=0}}$$
Let $A_{f}$ represent such a field  
$$A_{f}(x,\tau)=A(x,\tau){{\Big |}_{{\Delta\tau+f.\Delta x=0}\atop{\Delta\tau^2+\Delta x^2=0}}}{\dot{=}}A(x,\tau){\Big |}_{f},$$
with ${\Big |}_{f}$ being a short notation for the double constraint ${{\Big |}_{{d\tau+f.dx=0}\atop{\Delta\tau^2+\Delta x^2=0}}}$.
It would not make any sense defining such a field if this restriction (to a line) on its support could not be sustained during its time evolution governed by the standard wave equation in $(3+1)$-dimensions. It is remarkable, as we will see, that this makes a consistent field definition.

 \noindent The derivatives of $A_{f}(x,\tau),$ allowed by the constraint (\ref{f}), are the directional derivatives along $f,$ which with the use of (\ref{fmu}) we write as
\be
\label{fd}
\partial_{\mu}A(x,\tau)_{f}\Longrightarrow(\frac{\partial }{\partial x^{\mu}}+\frac{\partial \tau}{\partial x^{\mu}}\frac{\partial}{\partial \tau})A_{f}={\Big(}\frac{\partial }{\partial x^{\mu}}-f_{\mu}\frac{\partial}{\partial \tau}{\Big)}A_{f}{\dot =}\nabla_{\mu} A_{f},
\ee

With $\nabla$ replacing $\partial$ for taking care of the constraint (\ref{f}), $\tau$ can be treated as a fifth independent  coordinate, time-like and Lorentz invariant. The constraint (\ref{1}) and (\ref{dA0}) are used only afterwards then. We adopt this geometrical approach. It corresponds to embedding the physical spacetime in a $(3+2)$-manifold, as discussed in \cite{BJP}.  With the notation introduced in (\ref{fd}) the wave equation (\ref{we}) for a massless $A_{f}$ is rewritten as
\be
\label{wef}
\eta^{\mu\nu}\nabla_{\mu}\nabla_{\nu}A_{f}(x,\tau)=J(x,\tau),
\ee
or, explicitly 
$$(\eta^{\mu\nu}\partial_{\mu}\partial_{\nu}-2f^{\mu}\partial_{\mu})A_{f}(x,\tau)=J(x,\tau),$$
as $f^{2}=0$. $J$ is its source four-vector current. 

It can be solved by an f-Green's function,
\be
\label{sgf}
A_{f}(x,\tau_{x})=\int d^{4}yd\tau_{y}\; G_{f}(x-y,\tau_{x}-\tau_{y})\;J(y),
\ee
where the sub-indices specify the respective events $x$ and $y$, and $G_{f}(x-y,\tau_{x}-\tau_{y})$ being a solution of
\be
\label{gfe}
\eta^{\mu\nu}\nabla_{\mu}\nabla_{\nu}G_{f}(x-y,\tau_{x}-\tau_{y})=\delta^{4}(x-y)\delta(\tau_{x}-\tau_{y}).
\ee
This equation is solved \cite{ecwpd} to give:
\be
\label{pr9}
G_{f}(x,\tau)=\frac{1}{2}\theta(-b{\bar f}.x)\theta(b\tau)\delta(\tau+  f.x)=\frac{1}{2}\theta(-bf_{4}t)\theta(b\tau)\delta(\tau+  f.x),
\ee
where $b =\pm1,$ and $\theta (x)$ is the Heaviside function, $\theta(x\ge0)=1$ and $\theta(x<0)=0.$ It is clear from (\ref{pr9}) that $G_{f}(x,\tau)$ describes a point signal propagating with the four-velocity $f^{\mu}$ on a straight line $f$. The reduction of the field support from a lightcone to a lightcone generator reduces the discrete field to just a point in the phase space.

\section{The discrete Green's function.}
The properties of $G_{f}(x,\tau)$ will be properly discussed in \cite{ecwpd} but let us anticipate here the more relevant ones. 

The most obvious difference between $G_{f}(x,\tau)$ and the standard Green's function
\be
\l{sg} 
G(x,\tau)=\delta(t^2-r^2)=\frac{1}{r}[\delta(r-t)+\delta(r+t)],
\ee
 is the absence of singularity. The discrete field propagates without changing its amplitude. Such a so great difference between  two fields generated by a same source is closely associated to the distinct topologies of their respective supports. A hypercone is not a complete manifold in contradistinction to any of its generators.

$G_{f}(x,\tau)$ does not depend on ${\vec x}_{\hbox {\tiny T}}$, where the subscript ${\hbox {\tiny T}}$ stands for transversity with respect to ${\vec f}$: $${\vec f}.{\vec x}_{{\hbox {\tiny T}}}=0.$$ 
\be
\l{Gt}
\frac{\partial}{\partial x_{{\hbox{\tiny T}}}}G_{f}(x,\tau)=0.
\ee
The interaction propagated by $G_{f}$ is blind to ${\vec x}_{{\hbox{\tiny{T}}}}$. Anything at the transversal dimensions are not affected by and do not contribute to the interaction described by $G_{f}(x,\tau)$. Let us emphasise the importance of this. We are working with a field formalism defined on a (3+1)-spacetime but with respect to its dynamics the spacetime is effectively reduced to a (1+1)-spacetime, without any breaking of the explicit Lorentz covariance. This may be relevant on several fronts like, for example, theory of integrable systems, lightcone quantization,  Kaluza-Klein like theories, quark confinement, etc. 

As a consequence of eq. (\ref{Gt}) the eq. (\ref{wef}) may be replaced by
\be
\l{Gm}
\eta^{\mu\nu}\nabla_{\mu}\nabla_{\nu} G_{f}(x,\tau)=\delta(x_{{\hbox{\tiny L}}})\delta(t)\delta(\tau),
\ee
where the subscript ${\hbox {\tiny L}}$ stands for longitudinal with respect to ${\vec f}.$ It means that we can just ignore ${\vec x}_{\hbox {\tiny T}}$, omitting the $\delta^2({\vec x}_{\hbox {\tiny T}})$ on the RHS of (\ref{Gm}). ${\vec x}_{\hbox {\tiny T}}$ are ignorable coordinates.

The eq. (\ref{Gt}) supports our interpretation of $A_{f}$ as a single physical point object; its propagation does not depend on ${\vec x}_{\hbox {\tiny T}}$, or in other words, on anything outside $f$. Distinct abelian fields emitted by neighbouring point-sources do not see each other; each one of them can be treated as an independent single entity. 

Since we left the derivation of (\ref{pr9}) for the reference \cite{ecwpd} it may be instructive to verify that it is indeed a solution of (\ref{Gm}). 
As
\be
\l{tfx}
\nabla(\tau+f.x)=0,
\ee
\be
\eta^{\mu\nu}\nabla_{\mu}\nabla_{\nu} \theta(b\tau)=\nabla^{\nu}(-bf_{\nu}\delta(b\tau))=f^{2}\delta'(\tau)=o,
\ee
and 
\be
\eta^{\mu\nu}\nabla_{\mu}\nabla_{\nu}\theta(-b{\bar f}.x)=\nabla^{\nu}(b{\bar f}_{\nu}\delta(-b{\bar f}.x))=f^{2}\delta'({\bar f}.x)=o,
\ee
because $f^2=0$, 
we find that
$$
\eta^{\mu\nu}\nabla_{\mu}\nabla_{\nu} G_{f}(x)=\delta(\tau+f.x)\nabla\theta(b\tau).\nabla\theta(-b{\bar f}.x)=-f.{\bar f}\delta(\tau)\delta({\bar f}.x)\delta(\tau+f.x)=$$
$$=-(f_{4}^2+|{\vec f}|^2)\delta(\tau)\delta(f_{4}t-|{\vec f}|x_{\hbox{\tiny L}})\delta(f_{4}t+|{\vec f}|x_{\hbox{\tiny L}})=2f_{4}^{2}\delta(\tau)\delta(2f_{4}t)\delta(|{\vec f}|x_{\hbox{\tiny L}})=\delta(\tau)\delta(t)\delta(x_{\hbox{\tiny L}}).
$$

For $f^{\mu}=({\vec f}, f^{4})$, ${\bar f}$ is defined by ${\bar f}^{\mu}=(-{\vec f}, f^{4});$ $f$ and ${\bar f}$ are two opposing generators of a same lightcone; they are associated, respectively, to the $b=+1$ and to the $b=-1$ solutions and, therefore, to the processes of creation and annihilation of a discrete field. See the Figure 2. 

\hspace{-3cm}
\vglue-3cm

\begin{minipage}[]{7.0cm}
\hspace{-1cm}
\parbox[t]{5.0cm}{
\begin{figure}
\vglue-3cm
\epsfxsize=400pt
\epsfbox{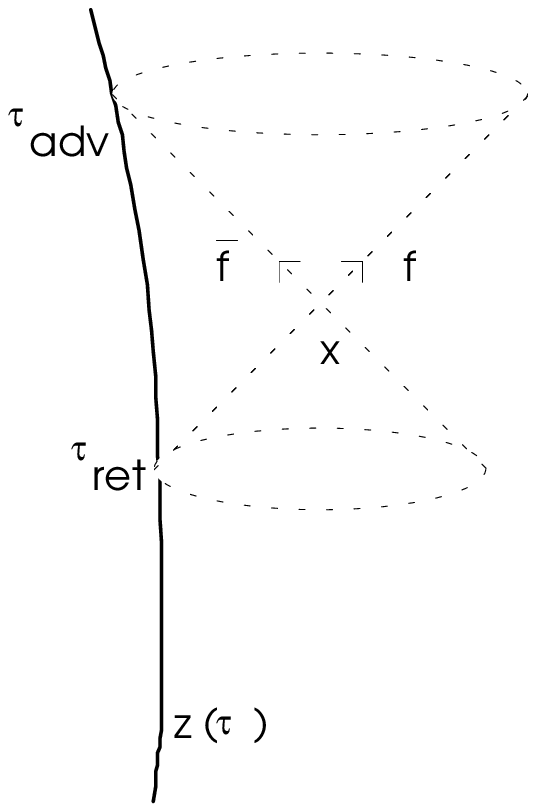}
\end{figure}}
\end{minipage}\hfill
\vglue-7cm

\hglue7.0cm

\begin{minipage}[]{7.0cm}\hglue12.0cm
\begin{figure}
\hglue7.0cm
\l{f3a}
\parbox[t]{5.0cm}{\vglue-6cm\hglue7.0cm{
\caption[Creation and annihilation of photons]{Creation and annihilation of particles: The $f$-Li\`enard-Wiechert solutions as creation and annihilation of particle-like fields. There are two classical photons at the point x: one, created at $\tau_{ret}$, has propagated to x on the lightcone generator f; the other one propagating on the lightcone generator ${\bar f}$ from x towards the electron worldline where it will be annihilated at $\tau_{adv}$.}}}
\end{figure}
\end{minipage}

\vglue-1cm

Observe in the Figure 2 that there is no backwards propagation in time implying that there is no advanced solution; the creation and the annihilation solutions are both retarded.
For $b=+1$ or $t>0$, $G_{f}(x,\tau)$ describes a point signal emitted by the electron  at $\tau_{ret},$ and that has propagated to $x$ along the fibre $f,$ of the future lightcone of $z(\tau_{ret})$;  for $b=-1$ or $t<0,$  $G_{f}(x,\tau)$ describes a point signal that is propagating  along the fibre $\bar{f}$ of the future lightcone of $x$ towards the point $z(\tau_{adv})$ where it will be absorbed (annihilated) by the electron. 
This is better visualized in the space diagram of the Figure 3 that shows the instantaneous ($t=const$) picture of the two solutions  moving from their respective sources. The only difference between the $(b=+1)$ and the $(b=-1)$ solutions is that $J$ is the source for the first and the sink for the second. Nothing else.
Observe the differences from the interpretation of the standard Li\`enard-Wiechert solutions. There is no advanced, causality violating solution here. $J$ is the source of the $f$-solution and a sink for the $\bar{f}$-solution. These two solutions correspond to creation and annihilation of discrete fields, exactly like well known processes of creation and annihilation of particles. 

\begin{center}{Gauge freedom and solutions from the homogeneous equation}\end{center}

Now we want to discuss the peculiar role of the boundary conditions and of the solutions from the homogeneous wave equation for fixing the discrete solution to (\ref{Gm}). Due to the lose link between the standard field $A(x)$ and its source one can always add arbitrary solution from its homogeneous equation. This is compatible with its gauge freedom and it is indeed necessary for attending some specific boundary conditions. For $A_{f}$ this is not so because the most general solution from its homogeneous equation \cite{ecwpd} is a generic function of $\tau+f.x$ which is trivially constant in the sense of (\ref{tfx}). So, the homogeneous equation admits only trivially constant discrete solutions. It is easy to understand why it is so. It describes interactions between point objects for which boundary conditions have no meaning. For the standard field the inhomogeneous solution is the field produced by the sources inside a given volume V while homogeneous solutions are fields produced\cite{Jackson} by external sources, i.e. the ones outside V. For a discrete field $A_{f},$ V is the line segment in $f$ between the two interaction points, and so any source must be on the boundaries (extremities) of this line segment $f$ but the charges there are the ones from $J$ and $J'$, the sources for the solutions from the inhomogeneous equation. Any other source beyond these points do not contribute to this $A_{f}$ although they may generate $A_{f}$'s in others segment of $f$. So, the solutions are necessarily the emitted and the absorbed-to-be solutions from the inhomogeneous equation.  There is no room left for any other source. See the Figure 3.


\begin{figure}
\epsfxsize=400pt
\epsfbox{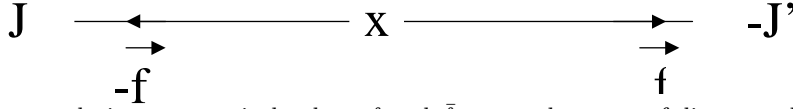}
\vglue-15cm
\caption[The two unique discrete solutions from the wave equation]{The two solutions, respectively along $f$ and  ${\bar f}$, spans the space of discrete solutions from the wave equation. The homogeneous eq. cannot have more than trivially constant solutions. Any other source on $f$, outside the line segment between $J$ and $J',$ are just outside the domain of definition of $A_{f}$ as a free field it is supposed to be, i.e. a field immediately after its creation event up to the event immediately before its annihilation.}
\end{figure}


So, eq. (\ref{pr9}) represents the unique (creation and absorption) solutions from eq. (\ref{Gm}). They form a basis for the space of all discrete solutions from the wave equation. This excludes the physical possibility of fields independent of sources, fields without sources, allowed in principle by this lose connection between an extended field and its source. The discrete field, in contradistinction to the standard one, is completely determined by its source. So. the discrete field has no gauge freedom; it is not a gauge field. This is important for understanding the meaning and the origin of gauge freedom of the extended field, whose discussion will be left for the Section IX.
\section{Retrieving the standard Green's function}

Here we prove the connection between $G(x)$ and $G(x,\tau)_{f}.$ First we define 
\be
\l{gg}
G(x,\tau){\dot =}\frac{1}{2\pi}\int d^4f\;\delta(f^2)G(x,\tau)_{f},
\ee
where 
\be
\l{df4}
d^{4}f=df_{4}\;|{\vec f}|^{2}\;d|{\vec f}|\;d^{2}\Omega_{f},
\ee
\be
\l{delta2}
\delta(f^{2})=\frac{1}{2|{\vec f}|}\{\delta(f^{4}-|{\vec f}|)+\delta(f^{4}+|{\vec f}|)\},
\ee
and then we show that $$G(x)=G(x,\tau)_{f}\delta(\tau).$$
The integration indicated in (\ref{gg}), according to (\ref{delta2}), is over the future ($f^4>0$) and the past ($f^4<0$) lightcones of $x$.  So in order to avoid double counting we must take just one solution from (\ref{pr9}), for example $b=+1$, and without the normalization factor 1/2: 
\be
\label{pr9a}
G_{f}(x,\tau)=\theta(-f_{4}t)\theta(\tau)\delta(\tau+  f.x).
\ee
We write $f.x=f_{4}t+r|{\vec f}|\cos\theta_{f},$ where the angle $\theta_{f}$ is defined by
\be
\l{theta}
{\vec f}.{\vec x}:= r|{\vec f}|\cos \theta_{f},
\ee
for a fixed ${\vec x}$. Then we have with (\ref{df4}) and (\ref{delta2}), and  after the integration on $f_{4}$,
\be
\l{gg2}
G(x,\tau)=\frac{\theta(\tau)}{4\pi}\int |{\vec f}| d|{\vec f}|d^2\Omega_{f}{\Big\{}\theta(-t)\delta{\big(}\tau+|{\vec 
f}|(r\cos\theta_{f}+t){\big)}+\theta(t)\delta{\big(}\tau+|{\vec f}|(r\cos\theta_{f}-t){\big)}{\Big \}}
\ee
and then

\be
\l{22}
G(x,\tau)=-\frac{\theta(\tau)}{2}\int_{-1}^{1}d\cos\theta_{f} \{\frac{\theta(-t)\tau}{(r\cos\theta_{f}+t)|r\cos\theta_{f}+t|}+\frac{\theta(t)\tau}{(r\cos\theta_{f}-t)|r\cos\theta_{f}-t|}\}.
\ee
The constraint (\ref{1}), $t^2=\tau^2+r^2$, implies that $|t|\ge r$. So,
\be
\l{tmr}
|r\cos\theta_{f}\pm t|=\frac{t}{|t|}(t\pm r\cos\theta_{f}),
\ee
and thus eq. (\ref{22}) becomes
\be
\l{qq}
G(x,\tau)=G(t,r,\tau)=-\frac{\theta(\tau)}{2}\int_{-1}^{1}d\cos\theta_{f}\;\frac{t\tau}{|t|}{\big\{}\frac{\theta(-t)}{(t+r\cos\theta_{f})^{2}}-\frac{\theta(t)}{(t-r\cos\theta_{f})^{2}}{\big\}}=
\ee
$$=\theta(\tau)\frac{t}{|t|}[\theta(t)-\theta(-t)]\int_{-1}^{1}
\frac{\tau}{(t+r\cos\theta_{f})^{2}}d\cos\theta_{f}.
$$
Therefore, considering that $$\frac{t}{|t|}[\theta(t)-\theta(-t)]=1,$$ we have
\be
\l{G}
G(x,\tau)=G(t^2-r^2,\tau)=\frac{\theta(\tau)\tau}{2(t^2-r^2)},
\ee

which, with the use of (\ref{1}) gives
\be
\l{qqq}
G(t^2-r^2,\tau){\Big |}_{\tau=0}=\cases{0,& for $|t|-r\ne0$;\cr
                   \infty,& for $|t|-r=0$.\cr}
\ee
The RHS of  eq. (\ref{G}) represents a spherical signal propagating with the velocity of light but it has not the correct dimension because it lacks the $(\tau=\sqrt{t^2-r^2}=0)$-constraint:$$\delta(\tau)=\delta(\sqrt{t^2-r^2}\;)=\frac{|\tau|}{r}[\delta(t-r)+\delta(t+r)],$$ which is considered only afterwards.
The appropriate Green's function is, therefore
\be
\l{Gd}
G(r,t)=G(t^2-r^2,\tau)\delta(\tau)=\frac{1}{2r}[\delta(t-r)+\delta(t+r)]=2\delta(t^2-r^2).
\ee

\begin{center}
\section{Retrieving the standard formalism}
\end{center}
Let us discuss the connections between the discrete field $A_{f}$ and $A$, the extended one. The eq. (\ref{gg}) implies that 
\be
\label{s1s}
A(x,\tau)=\frac{1}{2\pi}\int d^{4}f\;\delta(f^{2})A(x,\tau)_{f}.
\ee$A_{f}$ is not the intersection of $A$ with the fibre $f$ and
$A(x,\tau)$ does not represent a collection of all elements $A(x,\tau)_{f}$ from all possible fibres $f$'s. 
$A$ represents rather the smearing of $A_{f}$ over the lightcone. 
For the emitted $(f^{4}=|{\vec f}|)$ field in the source instantaneous rest frame at the emission time $(f^{4}=1)$ the equation (\ref{s1s}) can be written as
\be
\label{s}
A(x,\tau)=\frac{1}{4\pi}\int d^{2}\Omega_{f}A(x,\tau)_{f},
\ee
where the integral represents the sum over all directions of ${\vec f}$ on  a lightcone. $4\pi$, we see, is a normalization factor and (\ref{s}) is a particular case of
\be
\label{sss}
A(x,\tau)=\frac{\int_{\Omega_{f}} d^{2}\Omega_{f}A(x,\tau)_{f}}{\int_{\Omega_{f}} d^{2}\Omega_{f}}.
\ee
An integration over the $f$ degrees of freedom in (\ref{wef}) reproduces, with the use of (\ref{s1s}), the usual wave equation of the standard formalism, 
 as $$\int d^{2}\Omega_{f}f^{\mu}\partial_{\mu}\partial_{\tau}A_{f(x)}=0$$ because $A_{f}(x)$ is an even function of $f$, as we can see from (\ref{pr9}). The $f$-integration erases in the wave equation the effects of the constraint (\ref{f}) on the field. So, the standard extended formalism is retrieved from this discrete $f$-formalism with $A(x)$ as the average of $A_{f}(x)$, and (\ref{we}) as the average of (\ref{wef}), in the sense of (\ref{s1s}). 
\begin{center}
Physical interpretation.
\end{center}

The continuity of the extended field A is just an approximation. An electromagnetic wave, for example, is known to be made of a large number of photons, of which let us take $A_{f}$ as a classical representation. 
 This represents a drastic change in the meaning of $A(x)$. Being, in this new context, just an average field representation of the exchanged photon it can  produce good physical descriptions only at the measure of a large number of photons. This is the interpretation behind the equation (\ref{s1s}). It does not make much difference for most of the practical situations but it certainly fails for very low intensity light involving few photons. Imagine an extreme case of just one emitted photon, pictured as the fibre $f$ in the Figure 4, whereas $A(x)$ is the dotted circle. This is a very elucidative figure as it clearly shows that $A_{f}$ cannot be seem as the intersection of $A$ with the fibre $f$ and neither is the field $A$ a collection of $A_{f}$'s; it is obtained from the averaging (smearing) of each $A_{f}$ over the lightcone as defined by (\ref{s1s}). All informations about $f$ are irreversibly lost in this averaging process so that the reverse operation is not possible: $A_{f}$ cannot be obtained from $A$.  

\begin{minipage}[]{7.0cm}
\parbox[]{7.5cm}{
\begin{figure}
\vglue-6cm
\epsfxsize=400pt
\epsfbox{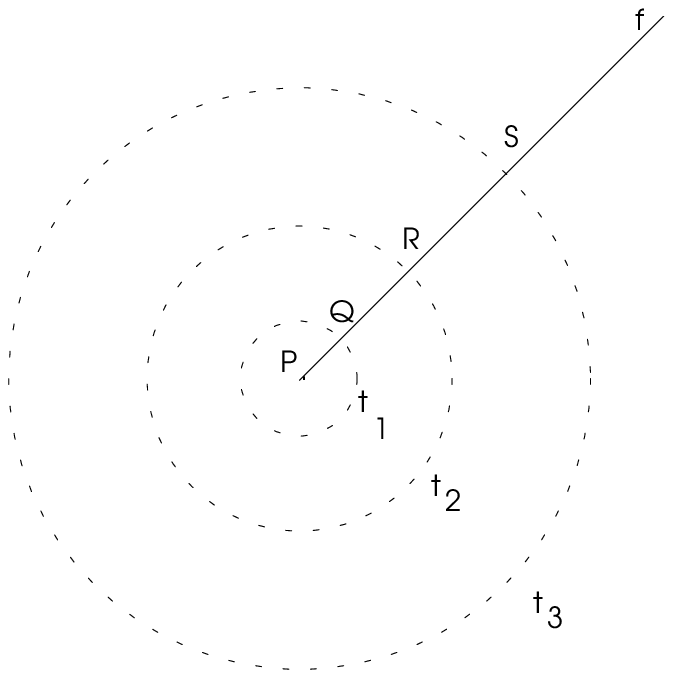}
\end{figure}}
\end{minipage}
\vglue-11cm
\hfill
\begin{minipage}[]{7.0cm}\hglue7.0cm
\begin{figure}[5cm]
\l{f2}
\parbox[]{5cm}{\vglue-1.5cm{
\caption[Extended and discrete fields]{The relationship between the fields $A_{f}$ and $A$. The three doted circles represent, at three instants of time, the field $A$ as an spherically symmetric signal emitted by a charge at the point P. The straight line PQRS\dots is the fibre $f,$ a null direction. The points Q, R, and S are a classical photon $A_{f}$ at three instants of time. In the case of a very low intensity light with just one photon the field $A$ transmit a false idea of isotropy.}}}
\end{figure}
\end{minipage}
\vglue1cm

Then $A(x),$  represented by the dotted circle, gives a false picture of isotropy while the true physical action (the photon) goes only along the $f$ null direction.  This is the origin  of problems (infinite self-energy, the Lorentz-Dirac as an equation of motion, etc) of classical electrodynamics at short distances \cite{Rorhlich,Jackson,Teitelboim,Rowe,Lozada}, discussed, in this context, in \cite{hep-th/9610028,ecwpd}.

\begin{center}
Physical and unphysical photons
\end{center}
\nopagebreak
There is more than just replacing (\ref{pr9}) in (\ref{gg}) in order to get 
(\ref{sg}). For a real photon $A_{f}$ the equation (\ref{theta}) makes no sense because $f$ is the photon four-vector velocity, it is co-linear to $x,\;$  $\;\theta_{f}=0.$ But the integration on $\cos\theta_{f}$ in (\ref{gg2}) is essential in order to get (\ref{sg}). We can understand (\ref{theta}) and (\ref{gg2}) in the following way: we have substituted the single physical photon $A_{f}$ by a continuous and isotropic distribution of fictitious photons $A_{f'}$ that are characterized by not having $f'$ co-linear to their direction of propagation $x$ ($x$ and $\tau$ are fixed in both sides of equation (\ref{gg2})). These fictitious fields $A_{f'}$ do not satisfy the condition (\ref{dA0}) and, consequently, as we will see in the following, they  are not transversal fields. So, in order to reproduce the Coulomb field, i.e. the $\frac{1}{r}$-potential, of the usual formalism based  on extended field  we necessarily need to introduce the unphysical longitudinal fields. The Coulomb continuous field $A$ necessarily contains spurious, non-physical,  gauge-violating discrete fields. For a quantum field this was known already by Dirac  and can be expected on the basis of some theorems of Quantum Field Theory \cite{Strochi}. These fictitious fields are the same trouble-maker ones that must be eliminated in a covariant quantization \cite{Gupta} of the extended field. The singularity presented by the average field $A(x)$, as shown on its Green's function (\ref{pr9}), is a consequence of this averaging process (\ref{gg}) in its definition. It comes from the lightcone support. This highlights the differences between $A_{f}(x,\tau)$ and $A(x)$ as also between their correspondent formalisms.\\

\begin{center}
The discrete and the Lienard-Wiechert solutions
\end{center}
 Assuming spherical symmetry around a single point charge, and neglecting  the constraint (\ref{dA0}), the $f$-integration on each $A_{f}$ over a lightcone produces the Lienard-Wiechert solutions. This does not imply, however that $A_{f}$ corresponds to the Lienard-Wiechert solutions because any standard solution of Maxwell's equations obtained from any charge and current distributions, and satisfying a boundary condition  can, in principle, be obtained from a linear combination of $A_{f}$'s in this averaging process.  The neglecting of the constraint (\ref{dA0}) and, therefore of the Lorentz condition, is a necessary condition for generating the standard formalism from this $f$-integration. 

One cannot do the same with the Lienard-Wiechert solutions; they cannot be used as a basis for the space of solutions of Maxwell's equations. The idea of a discrete field, an elementary unit of a field, cannot be extracted from them. That the Lienard-Wiechert solutions satisfy the Lorentz condition is a consequence of the structure of their source, $J(x)$. See the eqs. (\ref{s1s}) and the (\ref{BNA}) and (\ref{cc}) below.  Fields obeying other than the Lorentz condition can also be derived from an appropriate $f$-integration over $A_{f}$. Even static fields can be generated as linear combinations of discrete fields. They are seen as a certain limit of radiation fields. See the reference \cite{gr-qc/9801040} for the generation of the static Schwarzschild field of general relativity (the Einstein's equations become linear and reduces to the wave equation when applied to the discrete gravitational field), as an explicit example. It should not be necessary to say how this may be relevant to quantum gravity.

\section{Photon as a discrete field}
Let us now apply this discrete formalism to the electromagnetic field generated by a classical point spinless electron. We will take as an initial input just that the field is described by the wave equation (\ref{wef}), i.e. we just acknowledge its massless character ignoring any information about the Maxwell tensor structure (in other words, we will ignore eq. 3) and, therefore, any eventual connection between gauge symmetry and charge conservation: eq (\ref{wef}) just implies that\footnote{If we had (\ref{hs}),  we would have (\ref{iMa}) and would know that (\ref{we}) is a consequence of (\ref{L}) and, therefore, that the continuity equation was already contained in the Maxwell's equations (\ref{hM}) and (\ref{iM}).}.
\be
\l{BNA}
\Box\nabla.A_{f}=\nabla.J
\ee
For the discrete field, we will show now, everything else is already implicit in the spacetime structure, is a consequence of extended causality and Lorentz covariance.
We could (wrongly) assume that the Lorentz condition 
\be
\l{lg}
\nabla.A_{f}=0
\ee
is a consequence of the continuity equation
\be
\l{cc}
\nabla.J=0,
\ee
but that would be just apparently true. What we want to show is that (\ref{lg}) and (\ref{cc}) are both consequences of being discrete field relations.
The essence of the discrete formalism is that fields, sources and their interactions are discrete, point-like. The continuous image is just a macroscopic approximation. In this formalism where $\tau$ is treated as an independent fifth parameter, a definition of a four-vector current must carry an additional constraint expressing the causal relationship between two events $y$ and $z$. Its four-vector current is given by
\be
\l{J}
J^{\mu}(y,\tau_{y}=\tau_{z})= eV^{\mu}(\tau_{z})\delta^{3}({\vec y}-{\vec z})\delta(t_{y}-t_{z}),
\ee
where $z^{\mu}(\tau_{z}),$ is the electron worldline parameterized by its proper-time $\tau_{z}.$ In (\ref{J}) $\tau_{y}$ has to be equal to $\tau_{z}$ as a consequence of the Dirac deltas and of eq. (\ref{1}). For $b=+1$, that is, for the field emitted by J we have
\be
A_{f}(x,\tau_{x})=2e\int d^{5}y G_{f}(x-y)V^{\mu}(\tau_{y})\delta^{3}({\vec x}-{\vec y})\delta(t_{x}-t_{y}),
\ee
where the factor 2 accounts for a change of normalization with respect to (\ref{sgf}) as we are now excluding the annihilated photon (the integration over the future lightcone). Then,
\be
\l{Af1}
A_{f}(x,\tau_{x})=eV^{\mu}(\tau_{z})\theta(t_{x}-t_{z})\theta(\tau_{x}-\tau_{z}){\Big |}_{{\tau_{z}=\tau_{x}+f.(x-z)}},
\ee
which with $\Delta\tau=0$, is reduced to
\be
\l{A-f}
A_{f}(x,\tau_{x}=\tau_{z})=eV^{\mu}(\tau_{z})\theta(t_{x}-t_{z}){\Big |}_{{f.(x-z)=0}}
\ee

It is justified naming it discrete field because although being a field it is not null at just one space point at a given time. Its differentiability, in the sense of having space and time derivatives, is however assured by its dependence on $\tau$, a known continuous spacetime function. It is indeed a new concept of field, a very peculiar one, discrete and differentiable; it is just a finite point-like spacetime deformation running through a null direction. It is this discreteness in a field that allows the union of wave-like and particle-like properties in a same physical object, that implies finiteness and no spurious degree of freedom. 

Thus, the field $A_{f}$ is given, essentially, by the charge times its four-velocity at its retarded time. $\nabla\theta(t)$ and $\nabla\theta(\tau)$ do not contribute \cite{ecwpd} to $\nabla A_{f},$ except at $x=z(\tau),$ as a further consequence of the field constraints. So, for $\tau=0$ and $t>0$ we write just 
\be
\l{AfV}
A_{f}=eV{\Big |}_{f}
\ee
 and 
\be
\l{dAf}
\nabla_{\nu}A^{\mu}_{f}=\nabla_{\nu}(eV^{\mu}){\Big |}_{f}=-ef_{\nu}a^{\mu}{\Big |}_{f}.
\ee
The Lorentz gauge condition (\ref{lg}) and the continuity equation (\ref{cc}) are both consequences of (\ref{dA0}), i.e. of extended causality, and so they are automatically satisfied by the fields $A_{f}$ of (\ref{Af1}) and $J$ of (\ref{J}), respectively. They are just two identities. This is a very important point and deserves further elaboration. There is a causal link, a coherence, between a discrete field and its source that leads necessarily to (\ref{lg}) and (\ref{cc}). This link does not depend on the field tensorial or spinorial nature.  The extended causality constraint on a field leads to the constraint (\ref{dA0}) between its direction of propagation $f$ and the change in the state of motion of its source (sink) at the emission (absorption) time. The same goes for $\nabla.J=0,$ as $a.V=0$ is just a particular case of (\ref{dA0}) for $f=V$. There is a symmetry between a discrete field and its source that is partially hidden in classical electrodynamics by its oversimplified source representation. The classical representation (\ref{J}) for the electron current $J$ corresponds to a discrete-field current density. If we consider a vector field $j_{V}$ defined by 
\be
\l{j}
j^{\mu}_{V}=j^{\mu}{\Big |}_{V}=\int d^{4}y\;J^{\mu}(x-y)= eV^{\mu}(\tau){\Big |}_{V}
\ee
as describing the discrete current for this classical spinless electron the similarity between (\ref{AfV}) and (\ref{j}) becomes explicit: both are discrete fields as they have  hypercone generators for support, which are tangent, respectively to $f^{\mu}$ and to $V^{\mu}$. We have charge conservation regardless the Maxwell tensor antisymmetry which supposedly we don't know yet. Therefore charge conservation is also a consequence of (extended) causality and not of gauge symmetry. Although hiding the spinor character of the real electron, the electron discrete-current $j_{V}$ still shows the clear connection between (\ref{cc}) and (\ref{dA0}) because the physical content of (\ref{dA0}) is independent of the field tensorial or spinorial nature.

\begin{center}
The force field anti-symmetry
\end{center}
$A_{f}$, given by (\ref{AfV}), solution to (\ref{wef}) for a point charge $e$, is just an expression of the charge state of motion at the emission time, a instantaneous copy of (\ref{j}). State of motion or velocity is a relative or frame dependent concept and this suits well with $A_{f}$ being a potential; the force field, as well known, is associated to its gradient, that is, to the charge acceleration, which is an absolute concept.
 
We have started from eq. (\ref{wef}) assuming that we don't know anything about the force field $F_{f}$ except that it is massless and must somehow be associated to the gradient of $A_{f}$ and, therefore, in the case of $A_{f}$ being a vector field, to the irreducible components of a second-rank tensor
\be 
F^{\mu\nu}_{f}\Leftrightarrow-\nabla^{\nu}A_{f}^{\mu}=ef^{\nu}a^{\mu}{\big |}_{f},
\ee
defined by two four-vectors, the acceleration $a$ of its source at its emission time and $f$. Therefore, $F_{f}$ could in principle be a scalar or either a symmetric or an antisymmetric tensor. But $a$ and $f$ are not independent as they are constrained by (\ref{dA0}), which besides excluding the scalar component (the trace)  requires that ${\vec a}.{\vec f}=0,$ in the charge instantaneous rest-frame and this, as we show now, excludes the symmetric component. In other words, $F_{f}$ is a function not of $a$ but of $a_{\hbox{\tiny T}}$:
\be
\l{ft}
F_{f}\Leftrightarrow efa_{\hbox{\tiny T}}{\Big |}_{f}=ef(a-f\frac{{\vec a}.{\vec f}}{f_{4}^2}){\Big |}_{f},
\ee
 as $a_{\hbox{\tiny T}}:=a-f\frac{{\vec a}.{\vec f}}{f_{4}^2}.$ In this particular  frame the direction of propagation of $A_{f}$ is perpendicular to the electron instantaneous acceleration;  $A_{f}$ is a transversal field. But $ff\frac{{\vec a}.{\vec f}}{f_{4}^2}$ is not Lorentz  covariant (neither $a_{\hbox{\tiny T}}$) and there should be no privileged frame;  so it should not appear in the $F_{f}$ definition. Lorentz covariance requires an anti-symmetric tensor
\be
\l{FedA}
F^{f}_{\mu\nu}:=-e(f_{\mu}a_{\nu}-f_{\nu}a_{\mu}){\Big |}_{f}=\nabla_{\mu}A^{f}_{\nu}-\nabla_{\nu}A^{f}_{\mu}.
\ee
$F^{f}$ is given by the cross product of $f$ and $a$, so it is a transversal field.
This leads to the Maxwell's theory on a fibre $f$ embedded in the $(3+1)$-spacetime. 

This result (\ref{FedA}) cannot be obtained outside the extended causality context as the explicit dependence on $f$ is determinant. The constraint (\ref{dA0}) says that $F_{f}$ cannot depend on a longitudinal (with respect to $\vec{f}$) component of $a$. This, with Lorentz covariance, is enough to assure the anti-symmetry of $F_{f}$. For the extended field $A$ this is not true: the Lorentz condition (\ref{L}) excludes the scalar component of $F$ but not the symmetric one. It takes further constraints like the requirement of a transversal field, for example. This argument is extendable to non abelian interactions too.

\begin{center}
The $f$-electromagnetic field
\end{center}
The electromagnetic field of a classical photon $A_{f}$ is thus given by
\be
\l{cEf}
E_{f}^{i}=-F^{4i}_{f}{\Big |}_{f}=e(f^{4}a^{i}-f^{i}a^{4}){\Big |}_{f}=ef^{4}(a^{i}-f^{i}\frac{{\vec a}.{\vec f}}{(f_{4})^{2}}){\Big |}_{f}=ef^{4}a^{i}_{{\hbox{\tiny T}}}{\Big |}_{f},
\ee
\be
\l{Bf}
B^{i}_{f}=-\epsilon_{ijk}F^{jk}_{f}{\Big |}_{f}=e\epsilon_{ijk}f^{j}a^{k}{\Big |}_{f}=e\epsilon_{ijk}f^{j}a^{k}_{{\hbox{\tiny T}}}{\Big |}_{f}.
\ee
 
${\vec E}_{f},\;{\vec a}$ and ${\vec f}$ belong to a same plane, which is, by definition orthogonal to ${\vec B}_{f};\;$ $\;{\vec E}_{f}$ and ${\vec B}_{f}$ do not depend on ${\vec a}_{\hbox{\tiny L}},$ only on ${\vec a}_{\hbox{\tiny T}}.$  ${\vec E}_{f}$ is parallel to $a_{{\hbox{\tiny T}}}$. So there is no unphysical longitudinal photon i.e.  solution with longitudinal field:
${\vec E}_{f},\;{\vec B}_{f}$ and ${\vec f}$ form a triad of orthogonal vectors: ${\vec E}_{f}.{\vec B}_{f}=0,$\\
The $f-$Poynting vector ${\vec S}{_f}$ is defined by 
\be
\l{pv}
{\vec S}_{f}={\vec E}_{f}\times{\vec B}_{f}= e^{2}f^{4}{\vec a}_{{\hbox{\tiny T}}}\times({\vec f}\times{\vec a}){\Big |}_{f}=e^{2}f^{4}({\vec a}_{{\hbox{\tiny T}}})^{2}{\vec f}\;{\Big |}_{f}=e^{2}f^{4} a^{2}{\vec f}\;{\Big |}_{f},
\ee
with $a^{2}:=a^{\mu}a_{\mu}$ and we have used (\ref{a4}) to write 
\be
({\vec a}_{{\hbox{\tiny T}}})^{2}{\Big|}_{f}=({\vec a}-{\vec f}\frac{{\vec a}.{\vec f}}{f_{4}})^2{\Big|}_{f}=\{{\vec a}^{2}-(\frac{{\vec a}.{\vec f}}{f_{4}})^2\}{\Big|}_{f}= a^{2}{\Big|}_{f}
\ee
Having $F_{f}$ we  can derive its energy-momentum tensor as usual. It is given by 
\be
\l{thet}
\Theta_{f}^{\mu\nu}=F_{f}^{\mu\alpha}F_{f}^{\beta\nu}\eta_{\alpha\beta}+\frac{\eta^{\mu\nu}}{4}F_{f}^{\alpha\beta}F^{f}_{\alpha\beta}
\ee
 which after (\ref{FedA}) and (\ref{dA0}) is reduced to
 \be
\Theta_{f}^{\mu\nu}=-e^{2}f^{\mu}f^{\nu}a^{2}{\Big |}_{f}.
\ee
$\Theta_{f}$ is finite and represents the energy-momentum 
of $A_{f}$, a classical photon, a point object changelessly propagating along $f$.   $\nabla_{\nu}\Theta_{f}^{\mu\nu}=0$ is a  consequence of (\ref{dA0}); the energy-momentum content of $A_{f}$ is everywhere conserved. The Larmor theorem can be obtained from (\ref{thet}) as an average flux in the sense of (\ref{gg}), with the neglecting of (\ref{dA0}). $F_{f}$, but not $A_{f}$, always describes a classical photon. As $A_{f}$ reflects the state of motion of its source, it does not necessarily means an interaction, a classical photon, which requires a change in the source state of motion, an acceleration. 
If $z(\tau_{ret})$ is not a vertex, $A^{\mu}(x)$ does not describe a real classical photon; its Maxwell tensor is null. $F_{f}^{\mu\nu}\equiv\;0.$ A photon is a sudden change in the electron state of motion. If there is no change there is no photon and $A^{\mu}(x)$ is just a pure gauge field.\\

\vglue-0.5cm 
\begin{minipage}[]{5.0cm}\hglue-2.0cm
\parbox[]{5.0cm}{
\begin{figure}
\epsfxsize=200pt
\epsfbox{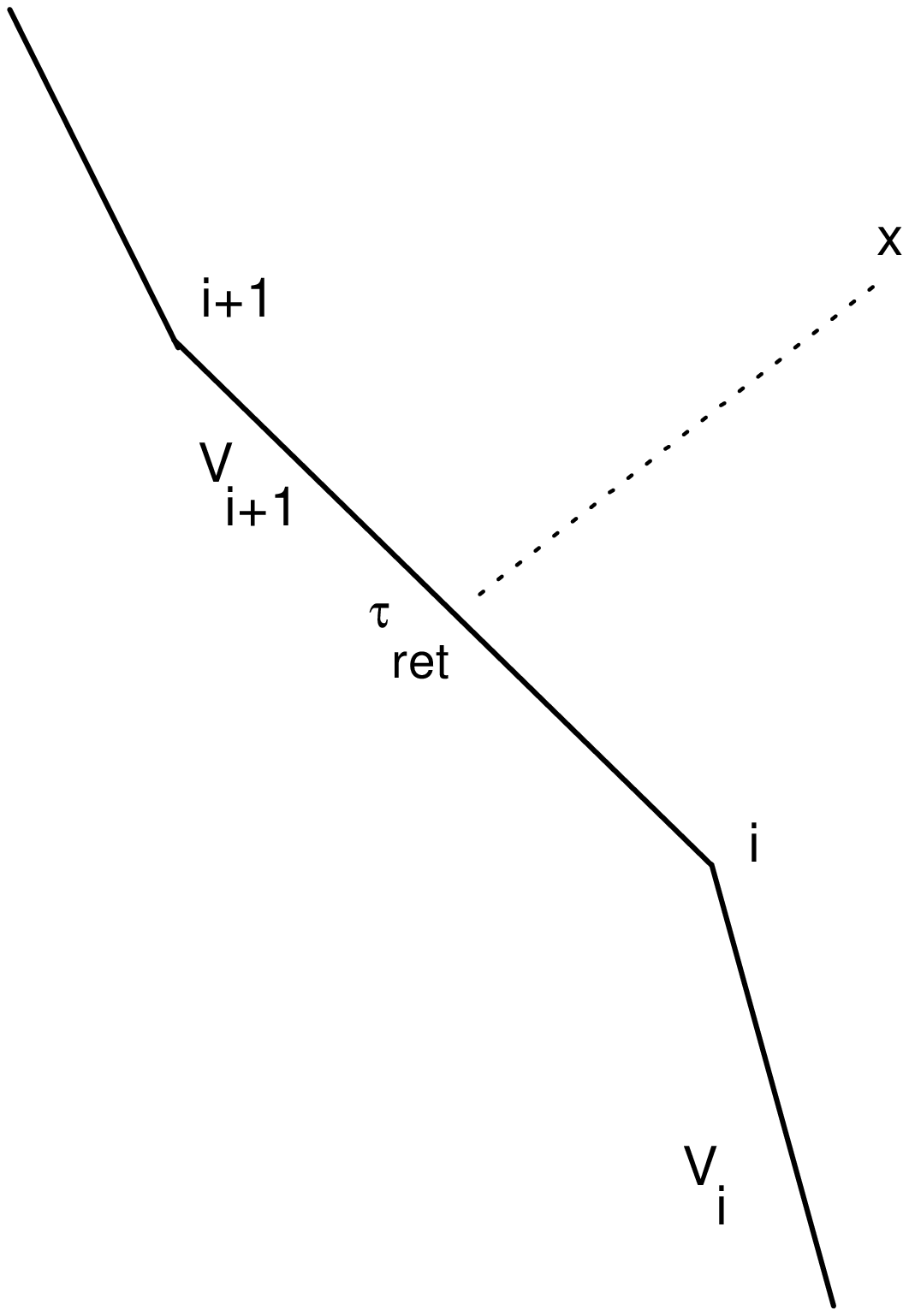}
\end{figure}}
\end{minipage}\hfill
\begin{minipage}[]{7.0cm}\hglue7.0cm
\begin{figure}
\l{18}
\parbox[b]{7.0cm}
{\vglue-7cm \parbox[t]{7.0cm}{\vglue-1cm \caption[Electron with a polygonal worldline.]{Classical photon from a polygonal-worldline electron. The dashed line is a null line. The past lightcone of $x$ picks just one event on the electron worldline. $A_{f}(x)$ is a real classical photon if this event is a vertex,  otherwise it is just a pure gauge field with $F_{f}^{\mu\nu}=0$.}
}}
\end{figure}
\end{minipage}\\ \mbox{}
\vglue-2cm

\section{Gauge freedom, singularity and duality}

$A_{f}$, regardless the $F_{f}$ antisymmetry, is completely determined by the state of motion of its source, and does not allow any non-trivial gauge freedom. It is not a gauge field. Extended causality, or equivalently, the concept of a field defined with support on a straight line, leads  to this unique solution (\ref{AfV}) for the wave equation, a field without any spurious degree of freedom. The Lorentz gauge condition for it, as a consequence of (\ref{dA0}), becomes an identity,
and one can understand why there is no symmetric field describing fundamental interactions. The important point here is that $A_{f}$ of eq. (\ref{A-f})  is not just a solution from the wave equation for an emitted radiation. It is the only solution for a given point source\footnote{Any linear combination of emitted ($t>0$) and absorbed ($t<0$) solutions is also a solution but they cannot be interpreted as an elementary physical entity.}. Any accelerated charge emits this same field $A_{f}$ and any extended field is a superposition of $A_{f}$'s. It is a universal field, in the same sense of an elementary particle, like a photon or an electron, is.

\noindent A remarkable point in the passage (\ref{s1s}) from a discrete to an extended field is the surging of gauge symmetry; even the Li\`enard-Wiechert solution, that satisfies (\ref{L}), has a residual gauge freedom. In $G$, $A$ and $F$  all information implicit with $f$ is lost with the $f$-integration. A generic solution $A$ is only indirectly linked to the state of motion of its sources; it can have many sources at once, even a continuous one or no source at all (solutions from homogeneous equation). In other words, with the $f$-integration the field $A_{f}$ becomes $A$, reduces the extended causality to just local causality, and acquires gauge freedom. The coherence between $A_{f}$ and its source at its emission time (expressed through $f$ as a one-to-one link between a field-event and a source-event) is lost. With $A(x)$ the fibre $f$ is replaced by the lightcone and a point-charge event is linked to an infinity of field events.

 \noindent Another remarkable point is that $A_{f}$ and $G_{f}$ have no singularity in contradistinction to their extended counterparts. This is a consequence of their distinct supports, a straight line and a lightcone, respectively. So, contrary to an old lore, the Coulomb (as also \cite{gr-qc/9801040} the Schwarszchild) singularity is not a consequence of a point-like source but just a reflex of the lightcone vertex \cite{BJP}. The field singularity appears with (\ref{s1s}), the integration over the lightcone, as a consequence of the field being defined as an effective average. This is saying again that the infinity in $A$ is introduced with (\ref{s}) and explains \cite{hep-th/9610028} the violation of causality in the Li\`enard-Wiechert advanced solution and the causality problems with the Lorentz-Dirac equation.

\begin{center}
A bridge to the quantum domain
\end{center}

Most of the difficulties of standard field theories comes from their use of field as an extended object, that is, with an infinite number of degrees of freedom. A discrete field means a rescue from that. It merges the ideas of field and of discreteness as an answer to the wave-particle duality. There is not a necessary contradiction or incompatibility between quantum physics and discrete fields. Conceptually, a discrete field is just a mathematical point in phase space (that is why it has a finite number of degrees of freedom) and this,  on the other hand, implies that its determination is fundamentally restricted to experimental uncertainties, which brings the connection to the (interpretation of the) Heisenberg's principle and to the field representation of a free particle. Both are fundamental points in the physical interpretation of quantum theory. Their close connections to the discrete-field concept will be discussed elsewhere.  A particle, in contradistinction to a discrete field, does not have wave properties (interference, etc) and cannot be made compatible with quantum mechanics. In general a photo detector is sensitive to the photon electric field and it cannot distinguish the fields of two or more discrete fields separated by a time interval smaller than its discriminating-time window (blind time-gap). It answers to the total electric field, say $\vec E_{T}=\vec E_{1}+\vec E_{2}$, and the amplitude of the total electric field $\vec E_{T}$ will depend on the phase difference between $\vec E_{1}$ and $\vec E_{2}$, exactly like in the theory of standard fields. The point is that it is dealing with a field, a pointlike field, but not a particle. In order to have interference we need fields, not necessarily waves nor extended fields. 
\section{The photon angular distribution}

The condition (\ref{dA0}) restricts the possible direction of emission of a classical photon by an accelerated charge: ${\vec f}$ is orthogonal to ${\vec a}$ in the charge instantaneous rest-frame. The photon is emitted in a plane orthogonal to ${\vec a}$. This is a maximally conceivable restriction as we are working with an (isotropic) spinless, point-particle model for the electron moving in a continuously smooth worldline.  A further final restriction must come only after considering the electron spin. 

In a laboratory  $\vec{a}$ is known only up to an experimental uncertainty $\Delta \vec{a}$. See the Figure 5. It is assumed here that we are dealing with an emitted classical photon, that is the case when $a$ determines $f$. In the inverse process of an absorbed classical photon, $f$ determines $a$, of course.
\vglue-3cm 
\begin{minipage}[]{5.0cm}\hglue-2.0cm
\parbox[]{5.0cm}{
\begin{figure}
\epsfxsize=400pt
\epsfbox{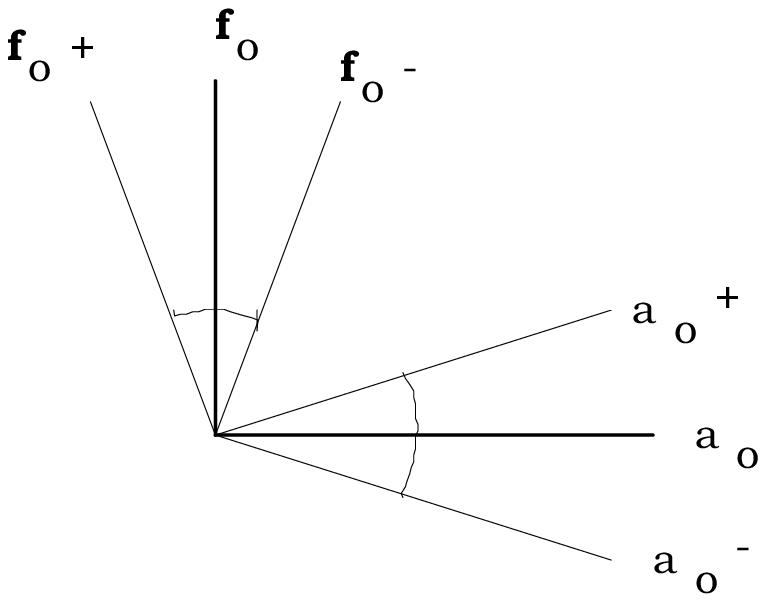}
\end{figure}}
\end{minipage}\hfill
\begin{minipage}[]{7.0cm}\hglue7.0cm
\begin{figure}
\l{14}
\parbox[b]{7.0cm}
{\vglue-9cm \parbox[t]{7.0cm}{\vglue-4cm \caption[Experimental uncertainty in the acceleration.]{$\vec{a}$, the space part of the electron acceleration, is known up to an experimental uncertainty $\Delta \vec{a}$.  Then there corresponds an indetermination $\Delta {\vec f}$, such that whichever be ${\vec a}$, inside its range $[{\vec a}_{0}-\Delta{\vec a},{\vec a_{0}}+\Delta{\vec a}],$ ${\vec f}$, in its range $[{\vec f}_{0}-\Delta{\vec f},{\vec f}_{0}+\Delta{\vec f}],$ will be in the plane orthogonal to ${\vec a}.$ Here ${\vec a}_{0}$ and ${\vec{f}}_{0}$ denote, respectively, the central or the most probable value of ${\vec{a}}$ and ${\vec{f}}$ in their respective ranges.}
}}
\end{figure}
\end{minipage}
\vglue-8cm
Then there corresponds, according to (\ref{af}), an indetermination $\Delta {\vec f}$, such that whichever be $\vec{a}$, inside its interval $[\vec{a}_{0}-\Delta\vec{a},\vec{a_{0}}+\Delta\vec{a}],$ ${\vec f}$, in its interval $[\vec{f}_{0}-\Delta\vec{f},\vec{f_{0}}+\Delta\vec{f}],$ will be in the plane orthogonal to $\vec{a}$, and that
\be
\vec{a}_{0}.\Delta\vec{f}+\vec{f}_{0}.\Delta\vec{a}=0,
\ee
\be
\vec{a}_{0}.\vec{f}_{0}=0,
\ee
are satisfied.
$\vec{a}_{0}$ and $\vec{f}_{0}$ denote the mean value of, respectively,  $\vec{a}$ and $\vec{f}$.\\ We need this for comparing the experimental data with the theory predictions. $a$ and $f$ in the theory always correspond to $a_{0}$ and $f_{0}$ inside their respective intervals in the experimental data.
\begin{center}
Straight line motion
\end{center}
The photon polarization, according to (\ref{cEf}) is determined by the charge instantaneous acceleration at the (retarded) emission time.
If the electron motion is such that ${\vec V}$ and ${\vec a}$ are collinear vectors, then with a boost along ${\vec V}$, see the Figure 6,  we have for the angle $\theta$ between ${\vec V}$ and ${\vec f}$:
\be
\tan\theta=\frac{\sin\theta'}{\gamma(\beta+\cos\theta')}=\frac{1}{\gamma\beta},
\ee
 as $\theta'=\pi/2$, or 
\be
\sin\theta=\sqrt{1-\beta^{2}},
\ee
where $\beta= |\frac{d{\vec z}}{dt_{z}}|$ and $\gamma=(1-\beta^{2})^{-\frac12}.$ So, a ultra-relativistic electron emits a photon in a very narrow cone about the ${\vec V}$ direction. The direction of photon emission coincides with the direction of maximum emission in the standard formalism \cite{Jackson,Ternov}. Again, the photon polarization, as we see from (\ref{cEf}), is along the electron instantaneous acceleration, regardless the electron velocity. In the Figure 5  ${\vec f}'$ and ${\vec a}'$ are, respectively, ${\vec f}$ and ${\vec a}$ on the electron instantaneous rest frame.
\vspace{-5.0cm}


\hglue-3.5cm
\parbox[]{7.5cm}{
\begin{figure}
\l{15}
\vglue-1.5cm
\hspace{1.0cm}
\epsfxsize=400pt
\epsfbox{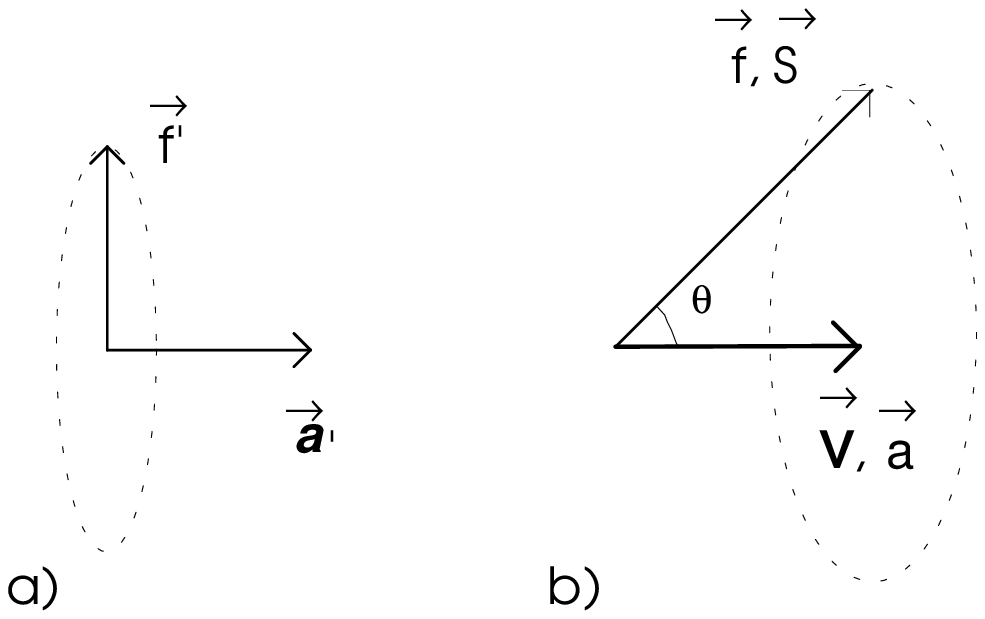}
\vglue-6cm
\end{figure}
\vglue-7cm
}
{}\\
\mbox{}


\vglue1cm
\hfill
\begin{minipage}[]{5.0cm}
\begin{figure}
\parbox[]{7.5cm}{\caption[Radiation from an electron in a straight-line motion.]{Radiation from an electron in a straight-line motion. a)The Lorentz condition requires that ${\vec f}'$ be orthogonal to the electron three-vector acceleration ${\vec a}'$ on its instantaneous rest frame: ${\vec a}'.{\vec f}'=0.$ b)In the laboratory frame the radiation is restricted to a cone of aperture $\theta$, with $sin\theta=\sqrt{1-\beta^{2}},$ $\quad{\vec\beta}=\frac{d{\vec z}}{d\tau},\; c=1.$}}
\end{figure}
\end{minipage}

\begin{center}
Circular motion
\end{center}
If the charge is in a circular motion then ${\vec a}.{\vec V}=0$, which from  $a.V=0$ implies also that $a^{4}=0$ and then, with (\ref{dA0}), that ${\vec a}.{\vec f}=0.$ Inverting (\ref{Bf}) we find that
\be
\l{fv1}
{\vec f}=\frac{{\vec a}\times{\vec B}_{f}+e{\vec a}{\vec f}.{\vec a}}{e{\vec a}.{\vec a}}=\frac{{\vec a}\times{\vec B}_{f}}{e{\vec a}.{\vec a}}.
\ee
\noindent Then we conclude from (\ref{fv1}) that ${\vec f}$, for a charge in a circular motion,  is orthogonal to the plane defined by ${\vec B}_{f}$  and ${\vec a}.$ But, in the absence of magnetic monopoles, ${\vec B}_{f}$ is always orthogonal to ${\vec V}$, its source's velocity, and so we have that the field synchrotron radiation is always emitted in the ${\vec V}$-direction. Again, this agrees with the direction of maximum emission of the standard formalism \cite{Jackson,Ternov}.

\section{Conclusions}
This work throws some light on the meaning and origin of gauge fields, their symmetries and singularities. If $A_{f}$ is accepted as the fundamental field and $A$ as just an effective average then the antisymmetry of $F$, the need of boundary conditions, the lose connection between $A$ and its sources, their gauge freedom and singularity can all be better understood. 
In a theory of discrete fields and sources each discrete field $A_{f}$ is causally linked  to the event of its creation by its point source. This link (or correlation) is represented by the null $f$-direction. $A_{f}$  being completely determined by the state of motion of its source has  no gauge freedom. 
When $A_{f}$ is integrated over $f$ in order to get $A$, this one-to-one link is lost and the field $A$ no more represents the state of motion of any particular point source as it is just an average field. The connection between the state of motion of a point source and a discrete solution of the inhomogeneous equation becomes lose as now any solution from the homogeneous equation can be added to it. $A_{f}$ is not a gauge field, a consequence of equation (\ref{dA0}) while $A(x)$ being a gauge field and its singularity are consequences of (\ref{s1s}).

With respect to the physical interpretation of the discrete electromagnetic field it would be a too easy solution to regard it as a mere mathematical artifact without

The richness of this approach lies on its promising applications to a wider and unifying quantum context. This picture replaces both particles and fields with a new kind of object: the discrete field, a particle-like field that is neither particle nor the usual extended field. $A_{f}$ as a field fits the wave properties of light and fits also, for being discrete, the particle properties like in the Compton and in the photo-electric effects. The discrete field, in this sense, is a representation of the wave-particle duality of quantum mechanics.

There are of course many other questions still to be answered. Our decision of applying the discrete formalism to a classical theory implied on accepting to work with its structural limitations which are mainly due to its emphasis on particle's trajectories and on a scalar representation for the sources. But even such a naive context has its merits others than just the simplicity. It is necessary now to distinguish what is just a consequence of the simple description adopted for the field sources from what may require a real quantum treatment. How far can one go with such a classical scheme? Where a legitimate quantum input must necessarily be added? What is a legitimate quantum input? What does it mean? These are just examples of valid questions that can be raised now.


\begin{thebibliography}{10}

\bibitem {hep-th/9610028} M. M. de Souza, J. of Phys. A: Math. Gen. 30 (1997)6565-6585.
\bibitem{ecwpd}in preparation.
\bibitem {gr-qc/9801040} M. M. de Souza, Robson N. Silveira,  Class. \& and Quantum Gravity, vol 16, 619(1999).
\bibitem {hep-th/9610145} M. M. de Souza, ``Classical Fields and the Quantum concept."hep-th/9610145.
\bibitem{Rorhlich}F. Rorhlich, ``Classical Charged Particles",  Reading, Mass. (1965).
\bibitem{Rorhlichp112}See the reference \cite{Rorhlich} at page 112.
\bibitem{Muta}N. Nakanish, ``Critical review of the theory of Quantum Electrodynamics" in  ``Quantum Electrodynamics", T. Kinoshita (ed), p.36, World Scientific, Singapure(1990).
\bibitem{Dixon} W. G. Dixon, Trans. Roy. Soc. Lond. A. 277, 59(1974) and the references therein.
\bibitem{Jackson} D. Jackson ``Classical Electrodynamics",2nd ed., chaps. 14 and 17,
John Wiley {\&} Sons, New York, NY(1975).
\bibitem{Teitelboim} C. Teitelboim; D. Villaroel; Ch.G.Van Weert, Rev. del
Nuovo Cim., vol 3, N.9,(1980).
\bibitem{Rowe} E.G.P.Rowe, Phys. Rev. D, 12, 1576(1975); 18, 3639(1978);  Nuovo Cim, B73, 226(1983).
\bibitem {Lozada} A. Lozada, J. Math. Phys., 30,1713(1989).
\bibitem {BJP} M. M. de Souza, Braz. J. of Phys., vol 28, n. 3, 250(1999).
\bibitem{tau}Fanchi, J. R., Found. Phys 23, 487(1993);Gaioli, F. H., Garcia-Alvarez, E. T., Int. J. Phys, 36, 2391(1997).
\bibitem{Strochi}Ferrari, R., Picasso, L. E., Strochi, F., Commun. Math. Phys. 35, 25(1974); F., Phys. Rev D 17, 2010(1978);Strochi, F., in Field Theory, Quantization and Statistical physics, E. Tirapegui (ed), 227-236, Reidel, The Netherlands, (1981).
\bibitem {Gupta} S. N. Gupta, Proc. Phys. Soc. A63, 681 (1950); K. Bleuler, Helv. Phys. Acta 23, 567 (1950).
\bibitem{Ternov} A.A. Sokolov, I. M. Ternov, {\it Radiation from relativistic electrons}, A.I.P, New York, N. Y, 1986.
\end{thebibliography}
\end{document}